# Mining Knowledge in Astrophysical Massive Data Sets

Massimo Brescia[a*], Giuseppe Longo[b], Fabio Pasian[c]

[a]*INAF – Osservatorio Astronomico di Capodimonte, Via Moiariello 16, 80131 Napoli, Italy*

[b]*Dipartimento di Fisica, Università degli Studi Federico II, Via Cintia 26, 80125 Napoli, Italy*

[c]*INAF – Osservatorio Astronomico di Trieste, Via Tiepolo 11, 34143 Trieste, Italy*

**Abstract**

**M**odern scientific data mainly consist of huge datasets gathered by a very large number of techniques and stored in very diversified and often incompatible data repositories. More in general, in the e-science environment, it is considered as a critical and urgent requirement to integrate services across distributed, heterogeneous, dynamic "virtual organizations" formed by different resources within a single enterprise. In the last decade, Astronomy has become an immensely data rich field due to the evolution of detectors (plates to digital to mosaics), telescopes and space instruments. The Virtual Observatory approach consists into the federation under common standards of all astronomical archives available worldwide, as well as data analysis, data mining and data exploration applications. The main drive behind such effort being that once the infrastructure will be completed, it will allow a new type of multi-wavelength, multi-epoch science which can only be barely imagined. Data Mining, or Knowledge Discovery in Databases, while being the main methodology to extract the scientific information contained in such MDS (Massive Data Sets), poses crucial problems since it has to orchestrate complex problems posed by transparent access to different computing environments, scalability of algorithms, reusability of resources, etc.
In the present paper we summarize the present status of the MDS in the Virtual Observatory and what is currently done and planned to bring advanced Data Mining methodologies in the case of the DAME (DAta Mining & Exploration) project.

## 1. Introduction

**M**odern scientific data mainly consist of huge datasets gathered by a very large number of techniques and stored in very diversified and often incompatible data repositories. More in general, in the e-science environment, it is considered as a critical and urgent requirement to integrate services across distributed, heterogeneous, dynamic "virtual organizations" formed by different resources within a single enterprise. As an example, the Astronomy and Astrophysics environment has become an immensely data rich field due to the evolution of detectors (plates to digital to mosaics), telescopes and space instruments. Different astrophysics areas share the same basic requirement: to be able to deal with massive and distributed datasets whereas possible integrated with services. A famous sentence states that *"While data doubles every year, useful information seems to be decreasing, creating a growing gap between the generation of data and our understanding of it"*.

---
[*] Corresponding author. Tel.: +39-081-5575-553; fax: +39-081-456710; e-mail: brescia@na.astro.it.

This new understanding includes knowing how to access, retrieve, analyze, mine and integrate data from disparate sources. But on the other hand, it is obvious that a scientist cannot and does not necessarily want to become an expert in the fields of ICT (Information & Communication Technology). The idea behind projects like DAME, described in this paper, is to provide a user friendly and standardized scientific gateway to easy the access, exploration, processing and understanding of massive data sets. In the field of astronomy, DAME represents a typical product of the emerging field of Astroinformatics.

Bioinformatics, Geoinformatics, Astroinformatics are growingly being recognized as the fourth leg of scientific research after experiment, theory and simulations. They arise from the pressing need to acquire the multi- disciplinary expertise which is needed to deal with the ongoing burst of data complexity and to perform data mining and exploration on MDS.

The data interoperability standardization approach in Astrophysics have been resulted in the Virtual Observatory. It consists into the federation under common standards of all astronomical archives available worldwide, as well as data analysis, data mining and data exploration applications. The main drive behind such effort being that once the infrastructure will be completed, it will allow a new type of multi-wavelength, multi-epoch science which can only be barely imagined.

**2. Massive Data Computing in Astrophysics**

An important part of the computing challenges in astronomy are related to the handling, processing and modeling of large quantities of data. In particular, processing of huge quantities of data (large detectors, mosaics, images with high time resolution) is typical of the optical and solar communities. The amount of computations needed to process the data is impressive, but often "embarrassingly parallel" since based on local operators, with a coarse grained level of parallelism. In such cases, the "memory footprint" of the applications allows to subdivide data in chunks, so as to fit the RAM available on the individual CPUs and to have each CPU to perform a single processing unit. Which kind of resources are necessary to tackle the processing and analysis of large quantities of data in astrophysics? In most cases "distributed supercomputers", i.e. a local cluster of PCs such as a Beowulf machine, or a set of PCs distributed over the network, can be an effective solution to the problem. In this case, the GRID paradigm can be considered to be an important step forward in the provision of the computing power needed to tackle the new challenges.

As for data, the concept of "distributed archives" is already familiar to the average astrophysicist. The leap forward in this case is to be able to organize the data repositories to allow efficient, transparent and uniform access: these are the basic goals of the Virtual Observatory (VO). In more than a sense, the VO is an extension of the classical Computational Grid; it fits perfectly the Data Grid concept, being based on storage and processing systems, and metadata and communications management services.

The Virtual Observatory (VO) is a paradigm to utilize multiple archives of astronomical data in an interoperating, integrated and logically centralized way, so to be able to "observe a virtual sky'" by position, wavelength and time. Not only data actually observed are included in this concept: theoretical and diagnostic can be included as well. VO represents a new type of a scientific organization for the era of information abundance:

- It is inherently *distributed,* and web-centric;
- It is fundamentally based on a *rapidly developing technology*;
- *It transcends the traditional boundaries* between different wavelength regimes, agency domains;
- It has an *unusually broad range of constituents* and interfaces;
- It is inherently *multidisciplinary*;

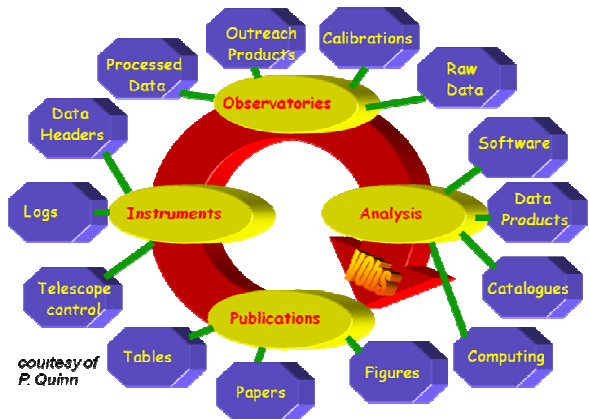

Fig. 1. The VO organization chart.

The International VO (cf. the IVO alliance or IVOA at http://ivoa.net) has opened a new frontier to astronomy. In fact, by making available at the click of a mouse an unprecedented wealth of data and by implementing common standards and procedures, the VO allows a new generation of scientists to tackle complex problems which were almost unthinkable only a decade ago [5]. Astronomers may now access a "virtual" parameter space of increasing complexity (hundreds or thousands features measured per object) and size (billions of objects).

## 3. A new perspective of distributed Data Mining

As underlined above, most of the implementation effort for the VO has concerned the storage, standardization and interoperability of the data together with the computational infrastructures. In particular it has focused on the realization of the low-level tools and on the definition of standards.

Our project DAME extends this fundamental target by integrating it in an infrastructure, joining service-oriented software and GRID resource-oriented hardware paradigms, including the implementation of advanced tools for MDS exploration, Soft Computing, Data Mining (DM) and Knowledge Discovery in Databases (KDD), [3]. Moreover, DM services often run synchronously. This means basically that they execute jobs during a single HTTP transaction. This might be considered useful and simple, but it does not scale well when it is applied to long-run tasks. Typical long-running activities are the following:

- any archive query traversing a massive DB table;

- a data-mining job running from a batch (sequential) queue;

- a pipeline workflow with several computing-intensive steps, applied sequentially for many (and massive) data sets.

In any of these cases, the system is stressed if the activity lasts longer than a few minutes and becomes unreasonably fragile if it lasts longer than a few hours. With synchronous operations, all the entities in the chain of command (client, workflow engine, broker, processing services) must remain up for the duration of the activity. If any component goes down or stops then the context of the activity is lost and the must be restarted. To overcome this limitation, one of the main DAME design strategies is to permit asynchronous access to the infrastructure tools, allowing running of activity jobs and processes outside the scope of any particular web-service operation and without depending on the user connection status.

The user, via client web applications, can asynchronously find out the state of the activity, has the possibility to keep track of his jobs by recovering related information (partial/complete results) without having the need to maintain open the communication socket. Moreover, the system is able to automatically perform a sort of garbage collection for cleaning up resources, swap areas and temporary system tools used during the activity run phase.

Furthermore, as it will be discussed in what follows, the DAME design takes into account the fact that the average scientists cannot and/or does not want to become an expert also in Computer Science. In most cases the r.m.s. scientist already possesses his own algorithms for data processing and analysis and has implemented private routines/pipelines to solve specific problems. These tools, however, often are not scalable to distributed computing environments. DAME aims at providing a user friendly web based tool to encapsulate own algorithm/procedure into the package, automatically formatted to follow internal programming standards.

The natural computing environment for MDS processing is a distributed infrastructure (GRID/CLOUD), but again, we need to define standards in the development of higher level interfaces, in order to:

- isolate end user (astronomer) from technical details of VO and GRID/CLOUD use and configuration;
- make it easier to combine existing services and resources into experiments;

Data Mining is usually conceived as an application (deterministic/stochastic algorithm) to extract unknown information from noisy data. This is basically true but in some way it is too much reductive with respect to the wide range covered by mining concept domains. More precisely, in DAME, data mining is intended as techniques of exploration on data, based on the combination between parameter space filtering, machine learning, soft computing techniques associated to a functional domain. The functional domain term arises from the conceptual taxonomy of research modes applicable on data. Dimensional reduction, classification, regression, prediction, clustering, filtering are example of functionalities belonging to the data mining conceptual domain, in which the various methods (models and algorithms) can be applied to explore data under a particular aspect, connected to the associated functionality scope.

The analytical methods based partially on statistical random choices (crossover/mutation) and on knowledge experience acquired (supervised and/or unsupervised adaptive learning) could realistically achieve the discovery of hidden laws behind focused phenomena, often based on nature laws, therefore the simplest.

During the R&D phase of our project, aimed at define and characterize rules, targets, ontology, semantics and syntax standards, the functional breakdown structure, outlined in Fig. 2, was derived. It provides a taxonomy between possible data exploration modes, made available by our infrastructure as data mining experiment typology (use case).

## 4. The DAME project overview

From the scientific point of view, the DAME project arises from the astrophysical domain, where the understanding of the universe beyond the Solar System is based on just a few information carriers: photons in several wavelengths, cosmic rays, neutrinos and gravitational waves.

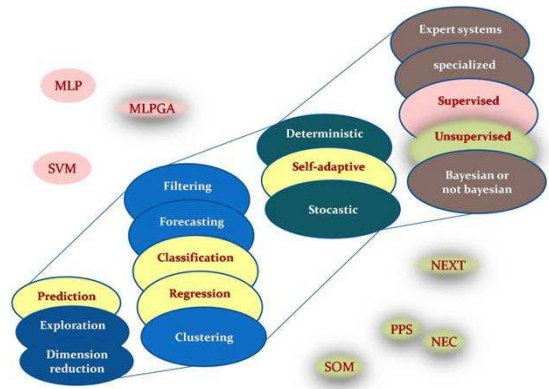

Fig. 2. Data Mining functional taxonomy

Each of these carriers has it peculiarities and weaknesses from the scientific point of view: they sample different energy ranges, endure different kinds and levels of interference during their cosmic journey (e.g. photons are absorbed while charged Cosmic Rays (CRs) are deflected by magnetic fields), sample different physical phenomena (e.g. thermal, non thermal and stimulated emission mechanisms), and require very different technologies for their detection. So far, the international community needs modern infrastructures for the exploitation of the ever increasing amount of data (of the order of PetaByte/year) produced by the new generation of telescopes and space borne instruments, as well as by numerical simulations of exploding complexity. Extending these requirements to other application fields, main goal of the project can be summarized in two items:

- The need of a "federation" of experimental data, by collecting them through several worldwide archives and by defining a series of standards for their formats and access protocols;
- The implementation of innovative computing tools for data exploration, mining and knowledge extraction, user-friendly, scalable and as much as possible asynchronous;

These topics require powerful, computationally distributed and adaptive tools able to explore, extract and correlate knowledge from multivariate massive datasets in a multi-dimensional parameter space. The latter results as a typical data mining requirement, dealing with many scientific, social and technological environments.

Concerning the specific astrophysical aspects, the problem, in fact, can be analytically expressed as follows:

Any observed (or simulated) datum defines a point (region) in a subset of $R^N$, such as:

- R.A. and DEC;
- time and λ;
- experimental setup (i.e. spatial and/or spectral resolution, limiting magnitude, brightness, etc.);
- fluxes;
- polarization;
- spectral response of the instrument;
- PSF;

Every time a new technology enlarges the parameter space or allows a better sampling of it, new discoveries are bound to take place.

So far, the scientific exploitation of a multi-band (D bands), multi-epoch (K epochs) universe implies to search for patterns and trends among N points in a DxK dimensional parameter space, where $N > 10^9$, $D >> 100$, $K > 10$.

The problem also requires a multi-disciplinary approach, covering aspects belonging to Astronomy, Physics, Biology, Information Technology, Artificial Intelligence, Engineering and Statistics environments. So far, experimental science has in practice become a three players game, made by:

- Scientists: theory, data, understanding, discoveries, structure, biases;
- Statisticians: evaluation of data, validation, analysis, dimensional reduction, models;
- Computer Scientists & Engineers: design and implementation of infrastructures, archives and databases, information retrieval, middleware, scalable tools, UML, OOP, metadata communication protocols

By taking into account both theoretical and conceptual domains, we designed an infrastructure based on the following skill features, [4]:

- DAME Suite composed of several software components which relies on a common infrastructure;
- Object Oriented Programming (OOP);
- Internal standards and protocols (VO, XML);
- Java language (almost generic for data mining models);
- User/Session DataBase Management System (MySQL);
- Web-based User I/O (Google Web Toolkit);
- Service-Oriented Web Application and Restful Web Service Technology, Servlet (Web Server applets);
- Plugin-based Modularity (easy to be integrated/modified) for data mining models;
- Hardware independent;
- Data conversion and manipulation support (ASCII, FITS, CSV, VOTable);

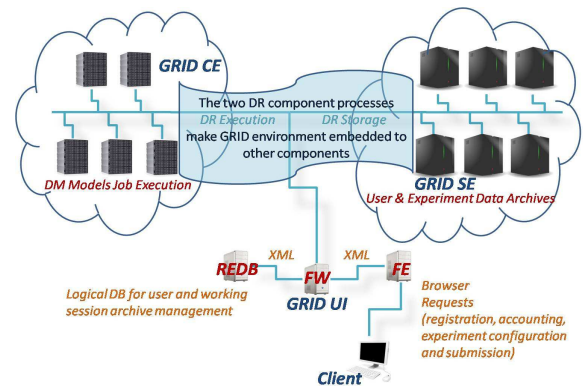

Fig. 3. DAME Infrastructure functional layout

To separate workflow functional requirements from their implementation issues, a library of methods called DRiver (DR) Management System has been provided (Fig. 3). The DR is the component used by the Framework (FW), core of the Suite, to manage the processing environment. It implements the low-level interface with computational environment, in order to permit the FW implementation, through the specific drivers, on different platforms (such as Stand-Alone or GRID). In other words, the DR is used to implement the proper access to the platform-dependent resources required by all the specific use cases and functionalities of the Suite. In order to avoid multiple deployment of these data on several components, it was decided to provide a file storage system (SE or Storage Element on GRID), hosting real data files, handled directly by the FW through the DR component methods. This component includes also a library of data file format (FITS, ASCII, CSV and VOTable) translating methods, used by the FW depending on the specific DM model data format supported. The DMM is the component that implements the DM (Data Mining) models and related wrappers code and their use cases. Its main features can be summarized as follows:

1. Implementation oriented to the functionalities (i.e. classes of functionalities, such as Classification and Regression);
2. Possibility of use functionalities with more than one model without duplicate code (Pattern Bridge as standard design pattern);
3. A common interface for all the models (by means of a specific class rendering typical data mining self-adaptive parameters, called DMMParams);

In the first release, the DMM implements MultiLayer Perceptron (MLP) with standard Back Propagation learning rule, MLPGA (MLP & Genetic Algorithms) and Support Vector Machine (SVM) as supervised models. All these models have a common data mining paradigm: the AI (Artificial Intelligence) technique as self-adaptive exploration methodology. The experiment that user can configure through our suite is in practice made of the choice between available functionalities and related DM model.
The models and algorithms embedded in the project have been already tested on astrophysical use cases, with successful results, as reported in [1] and [2].

## Acknowledgments

The DAME project (http://voneural.na.infn.it), run jointly by the Department of Physics of the University Federico II, INAF (National Institute of Astrophysics) Astronomical Observatory of Napoli, and the California Institute of Technology, is financed through grants from the *Italian Ministry of Foreign Affairs*, the *European projects VO-TECH and VO-AIDA (Astronomical Observatory of Trieste)* and by the USA - *National Science Foundation*. DAME makes use of distributed computing environments (e.g. the S.Co.P.E. - GRISU infrastructure) and matches the international IVOA standards and requirements. The authors thank all people involved in the DAME working group, whose contribution have made available the described technological and scientific products.